\documentclass{aastex} \usepackage[numberedappendix]{emulateapj5}
\shorttitle{Cepheid Distance to M33} \shortauthors{Lee et al.}

\begin{document}

\title{
DETERMINATION OF THE DISTANCE TO M33
BASED ON SINGLE EPOCH $I$-BAND $HST$ OBSERVATIONS OF CEPHEIDS$^1$}
\altaffiltext{1}{Based on observations with the NASA/ESA Hubble Space
Telescope obtained at the Space Telescope Science Institute, which is
operated by the Association of Universities for Research in Astronomy,
Incorporated, under NASA contract NAS5-26555.}

\author{Myung Gyoon Lee, Minsun Kim}
\affil{Astronomy Program, SEES, Seoul National University, Seoul, 151-742, Korea}
\email{mglee@astrog.snu.ac.kr, mskim@astro.snu.ac.kr}

\author{Ata Sarajedini}
\affil{Department of Astronomy, University of Florida, P.O. Box 112055, 
Gainesville, FL, 32611, USA} \email{ata@astro.ufl.edu}

\author{Doug Geisler, and Wolfgang Gieren}
\affil{Departamento de F{\'\i}sica, Grupo de Astronom{\'\i}a, Universidad de Concepci\'on,
	Casilla 160-C, Concepci\'on, Chile}
\email{doug@kukita.cfm.udec.cl, wgieren@coma.cfm.udec.cl}

\begin{abstract}
We have determined the distance to M33 using single epoch $I$-band observations of Cepheids
based on $HST/WFPC2$ images of five fields in M33.
Combining the $HST$ $I$-band photometry and the periods determined from the ground-based
observations (DIRECT) for 21 Cepheids with $\log P >0.8$ 
in the sample of 32 Cepheids,
we derive a distance modulus of
$(m-M)_0=24.52\pm0.14$(random)$\pm0.13$(systematic) 
%$(m-M)_0=24.41\pm0.13$(random)$\pm0.12$(systematic)
for an adopted total 
reddening of M33, $E(B-V)=0.20\pm0.04$ ($E(V-I)=0.27\pm0.05$) given by
\citet{fre01}, 
the reddening to the LMC, $E(B-V)=0.10$,
and the distance to the LMC, $(m-M)_0=18.50$.
If the total reddening to M33 of $E(B-V)=0.10\pm0.09$ given by \citet{fre91} is used,
the Cepheid distance modulus based on the $I$-band photometry will be increased by 0.20. 
Metallicity effect on the Cepheid distance to M33 is 
estimated to be small, $\delta (m-M)_Z=0.01$ to $0.06$, which leads to
$(m-M)_0=24.53$ to 24.58 after this metallicity effect correction.
%The source of the largest systematic error (0.37) is the total $I$-band extinction.
Using the Wesenheit $W_{I}$, an extinction-free parameter,
we derive a similar value, $(m-M)_0=24.52\pm0.15$(random)$\pm0.11$(systematic).
%$(m-M)_0=24.49\pm0.13$(random)$\pm0.10$(systematic).
These results are in reasonable agreement with those based on the ground-based 
multi-epoch $BVRI$ observations of brighter Cepheids in M33, 
and are $\approx 0.3$ smaller than those based on the tip of the red giant branch
and the red clump.
It is needed to estimate better the reddening to Cepheids in M33.
\end{abstract}


\section{INTRODUCTION}
The Local Group spiral galaxy M33 is one of the primary calibrators for
secondary distance indicators including the Tully-Fisher relation.
Although it is a nearby bright galaxy and the Cepheids in it were discovered 
as early as the 1920's \citep{hub26},
it was only in the 1980's that reasonable estimates for the distance 
to M33 became available \citep{san83a,san83b,fre91}. 
%(\citep{fre91} and references therein).

M33 is close enough so that useful photometry of the bright Cepheids
can be obtained from ground-based observations.
Taking advantage of the excellent seeing at the Canada-France-Hawaii 
Telescope, \citet{fre91} determined the distance
to M33 using $BVRI$ photometry of 10 bright Cepheids, 
obtaining $(m-M)_0 = 24.64\pm0.09$
%and $E(B-V)=0.10\pm0.09$.
which is close to the median of the previous estimates using Cepheids,
$(m-M)_0=24.1$ to 24.8 
\citep{san83a,san83b,mad85,chr87,mou87} (see also \citet{van00}).
%\citet{san83b}; \citet{mad85};\citet{chr87};\citet{mou87}; see also %\citet{van00}).
Later \citet{fre01} revised \citet{fre91}'s value to
$(m-M)_0 = 24.62\pm0.10$, adopting slightly different period-luminosity
relations and the metallicity effect correction of 0.06 mag
based on $\delta (m-M)_0 / \delta Z = -0.2 \pm0.2$ mag dex$^{-1}$.
On the other hand, very recently \citet{mac01} discovered 251 Cepheids in M33 and
presented their $BVI$ light curves using the FLWO 1.2m telescope. 
However, crowding and blending problems for faint Cepheids are severe in the ground-based data.
Therefore, the periods of the Cepheids in nearby galaxies like M33 can be determined 
reasonably well from the
ground-based observations, but the photometry of these Cepheids is prone to errors
 due to the severe crowding and blending \citep{moc01}.

In this paper, we present a determination of the distance to M33 using single epoch
$I$-band photometry of known M33 Cepheids, based on deep Hubble Space Telescope
Wide Field Planetary Camera 2 ($HST/WFPC2$) images, taking
advantage of the high spatial resolution of $HST$ and the periods of the DIRECT Cepheids
obtained from ground-based CCD observations.

Single epoch $I$-band photometry of Cepheids with known periods 
is very efficient when used in determining
the distances to nearby galaxies %when applied to Cepheids with known periods
\citep{fre88,lee93a,lee93b}.
The amplitude ($\sim 0.5$ mag) of variability of Cepheids and extinction
at the $I$-band are much smaller than at shorter wavelengths (such as $B$) 
so that we can
estimate the distances reasonably well 
even from single epoch $I$-band observations of several Cepheids with known
periods.

\section{OBSERVATIONS AND REDUCTION}

We analyzed $HST/WFPC2$ data for five fields in M33 
obtained between 1995 November to 1997 June  
for \citet{sar98}'s cycle 5 program (GO-5914).
Each field was observed for four orbits, yielding
a total exposure time of 4800 seconds for $F555W (V)$ and 5200 seconds for $ F814W (I)$.
These data were obtained originally for the study of globular clusters in M33;
thus, a globular cluster is centered in each PC chip.

We have identified 32 Cepheids known from the DIRECT project in these $HST$ images, 
as shown in Figure 1
(There were three more Cepheids in our fields, but two of them were saturated and
one could not be identified, so we used only 32 Cepheids for this study).
Table 1 lists the information for the Cepheids used in the present study.
The periods of these Cepheids range from 4 days to 26 days.
We have classified the quality of the light curves given by \citet{mac01}
into five classes: 0 for very good, 1 for good, 2 for fair,
3 for unusually red color, and 4 for ambiguous identification, 
as shown in Table 1.
At the position of C49 shown in the DIRECT image, 
there are seen two stars separated by 0.525 arcsecond in the $HST$ image.
The magnitudes and colors of both stars 
($V=20.78$, $(V-I)=0.59$, and $V=21.93$, $(V-I)=1.14$) are within 
the range of those of known Cepheids so that we could not identify which of the two is
a Cepheid. Finally we decided not to use this object for the distance estimation.

The photometry of the stars in the images
has been obtained using the {\it multiphot} routine of the HSTphot package
which was designed for photometry of $HST/WFPC2$ data and
employs a library of Tiny Tim  point-spread-functions (PSFs) 
for the PSF fitting \citep{dol00a,dol00b}.
The {\it multiphot} routine gives the magnitudes
transformed to the standard system as well as instrumental magnitudes.
The standard $V$ and $(V-I)$ of the Cepheids as measured by 
the {\it multiphot} routine are listed in Table 1.
Formal errors of both $V$ and $(V-I)$ are smaller than 0.02 mag.
More details of the observations and data reduction are given in \citet{kim01}.

After phasing our data to the DIRECT data, we have compared our photometry with the
DIRECT photometry on the same phase and with the mean magnitudes of the Cepheids
given by the DIRECT project \citep{mac01}, which are listed in Table 1. 
It is found that the phase distribution of our $HST$ Cepheid data is random, showing
that they can be used for reliable distance estimation.
A comparison of our single epoch photometry with both sets of the DIRECT data,
%the mean magnitudes  of the same Cepheids given by the DIRECT project 
shows that $HST$ magnitudes are on average $\approx 0.2$ mag 
fainter than the DIRECT magnitudes;
the differences between the two sets of photometry are 
$\Delta V$ (HST - DIRECT) = $0.16$ with a standard deviation $\sigma=0.46$,
and $\Delta I$(HST -- DIRECT) = $0.23$ with a  standard deviation $\sigma=0.29$, as shown in Figure 2. 
There is little difference in the standard deviation between the comparisons with 
the mean magnitudes and phased magnitudes of the DIRECT data.
This difference between our data and the DIRECT data is most likely
 due to the crowding and blending effect in the ground-based data 
which leads to brighter magnitudes in the DIRECT photometry (see also \citet{moc01}). 
However, the crowding and blending has much less effect on the period determination for
Cepheids so that the Cepheid periods given by the DIRECT project are considered 
to be reliable. The reason for the difference between our photometry and Mould's photometry 
(see below) is not known but note that it goes
in the direction opposite to what one would expect from additional crowding
in the ground-based data.

%For foreground reddening correction, the COBE/IRAS extinction maps of  are used.
The foreground reddening values of all the regions in M33 
are as low as $E(B-V)=0.04$ \citep{sch98}. % as listed in Table 1 \citep{sch98}.
\citet{fre91} estimated the mean value of the total (foreground plus internal)
reddening for the M33 Cepheids
from $BVRI$ photometry of bright Cepheids, to be $E(B-V)=0.10\pm0.09$.
Later \citet{fre01} revised this estimate 
to a value twice larger but with a smaller error,
$E(V-I)=0.27\pm0.05$ ($E(B-V)=0.20\pm0.04$), by applying new period-luminosity
relations of the Cepheids to the same data. 
We adopted the latter in this study.
The extinction laws for $R_V=3.3$, $A_{I}=1.956E(B-V)$(=0.39) and $E(V-I)=1.35E(B-V)$, 
\citep{car89} are adopted in this study. 
%$A_{I}=1.45 E(V-I)= 1.95E(B-V)= 0.39$, $A_{V}=2.45E(V-I) = 0.66$ 

\section{RESULTS}

\subsection{Color-Magnitude Diagram}

Figure 3 displays the color-magnitude diagram of 32 Cepheids listed in Table 1, 
where the field stars in one region (H38 region) are also included 
to illustrate the field stellar population.
%as an example of field stars. 
In Figure 3, we also plot for comparison
the mean magnitudes and colors of the same Cepheids given by the DIRECT
project \citep{mac01} and those of bright Cepheids given by \citet{fre91}.
%All the Cepheids are found to be located in the typical instability strip.
The magnitudes of the Cepheids range from $I=21.1$ mag to 18.5 mag
($V=22.0$ mag to 20.1 mag), reaching much fainter than those of the Cepheids used in
\citet{fre91}'s study (as shown by the open squares in Figure 3).
One M33 Cepheid (C150) with the reddest color ($V-I=1.73$ 
both from the $HST$ and DIRECT photometry) 
shows an extremely red color even compared with the Cepheids in other galaxies
\citep{fer00}.
The range of Cepheid colors obtained in this study is similar to that of the 
DIRECT project, 
although our photometry is based on just single epoch observations, 
while the DIRECT photometry represents mean values based on multi-epoch 
data ($>100$ photometric points taken over  about 40 nights). 

In Figure 4, M33 Cepheids are compared with those in the LMC \citep{uda99a}
and the $HST$ $H_0$ Key Project galaxies \citep{fer00, fre01}in the 
the color-magnitude diagrams.
%The LMC Cepheid data are from the OGLE data \citep{uda99b}.
In the case of the LMC for which the extinction $E(B-V)=0.1$ and the distance
 modulus $(m-M)_{0}=18.5$ are adopted, 
first overtone mode Cepheids (triangles) as well as the fundamental mode 
Cepheids (open circles) are plotted.
For  the $HST$ $H_0$ Key Project galaxies,
extinction values and distance moduli given by \citet{fre01} are adopted.
Figure 4 shows that M33 Cepheids used in this study are located
within the instabililty strip roughly defined by the Cepheids in other galaxies.
M33 Cepheids go fainter than those in the $HST$ $H_0$ Key Project galaxies,
and show a larger range of color compared with the LMC Cepheids.

\subsection{Cepheid Distance}

Figure 5 displays the $I- \log P$ relation of the Cepheids in M33 
based on this study %(filled circles, open circles, and an asterisk) 
and other studies:
mean magnitudes of bright Cepheids by \citet{fre91} (open squares),
mean magnitudes from the DIRECT project for the same Cepheids 
by \citet{mac01} (crosses), and
single epoch photometry of other Cepheids by \citet{mou87} (open triangles).
Several features of note are seen in Figure 5.
First, our $HST$ photometry of M33 Cepheids shows a tight correlation
between period and luminosity,
although our data are based only on single epoch observations.
Second, our photometry is on average fainter than the DIRECT photometry and brighter than
Mould (1987)'s photometry for a given period.
Third, our photometry is on average similar to Freedman et al. (1991)'s
photometry for a given period.
Fourth, the scatter along the period-luminosity relation is smaller in our 
photometry ($\sigma = 0.25$) than
that of the ground-based data ($\sigma = 0.28$ for the DIRECT data, 
$\sigma = 0.27$ for the Freedman et al.'s data, and $\sigma = 0.45$ for Mould's data).

For distance estimation, we have used the calibration of the 
$M_I - \log P$ relation
for Cepheids given by \citet{fre01}: %and \citet{uda99a}, %\citet{mad91}, 
$M_I = -2.962 \log P -1.942$ with $\sigma = 0.11$.
The zero point in this calibration is based on the LMC distance modulus of
$(m-M)_0=18.50$ and reddening $E(B-V)=0.10$.
The slope in this calibration is based on the results of \citet{uda99a}
and is slightly flatter than that given by \citet{mad91}, 
$M_I = -3.06 \log P -1.81$ with $\sigma = 0.18$.
The slope in the adopted calibration is similar to those based on newer studies \citep{gro00a,gro00b}.

There is a possibility that there may be included some first overtone
Cepheids as well as the fundamental mode Cepheids at the short periods.
As a matter of fact 
we could classify C29 and C35 which are the brightest among the Cepheids
with $\log P <0.8$ into first overtone Cepheids from the shape
of the light curves, noting that the light curves of the first overtone Cepheids
are more sinusoidal than the fundamental mode Cepheids \citep{man92}.
The LMC Cepheid data show that the longest period of the first overtone
Cepheids is about 6 days \citep{uda99a}.
Therefore we decided to use the Cepheids with $\log P > 0.8$ for
distance determination. 

Fitting the $I- \log P$ relation to 21 Cepheids (classes 0, 1 and 2)
with $\log P > 0.8$ in M33, we obtain
a value for the distance modulus, $(m-M)_I=24.91$ with $\sigma$(fit)$=0.25$ mag.
%$(m-M)_I=24.85$ with $\sigma$(fit)$=0.27$ mag. M91
The uncertainty corresponding to this fitting error is $\sigma$(fit)$/\sqrt{N}=0.05$ mag.
On the other hand, the error in the distance modulus associated with a single
$I$-band observation of one Cepheid of known period is 0.30 mag which leads 
to 0.065 mag for 21 Cepheids in this study, following the description in \citet{fre88}. 
These two types of errors are comparable so we adopt 0.07 as the error of
the $I- \log P$ fitting.
From this we derive an extinction-corrected distance modulus 
$(m-M)_0=24.52\pm0.14$(random)$\pm0.13$(systematic),
%$(m-M)_0=24.65\pm0.13$(random)$\pm0.20$(systematic),
considering extinction and
other error sources  as listed in Table 2 (following also \citet{mou00}).
If we use 18 good Cepheids with classes 0 and 1, 
we obtain very similar results, $(m-M)_0=24.50\pm0.14$(random)$\pm0.13$(systematic)
%$(m-M)_0=24.60\pm0.13$(random)$\pm0.20$(systematic).
%The source of the largest systematic error is that associated with the total
%$I$-band extinction given by \citet{fre91}.

In addition, we have used $W$ (the Wesenheit parameter) 
% which is an extinction-free parameter) 
for distance estimation (see \citet{fre91} ).
$W$ is a representative magnitude which is defined to be extinction-free:
$W_V=V-R_V (B-V)$ and $W_I=I-R_I (V-I)$ where $R_V$ and $R_I$ are the ratio of 
total-to-selective absorption. 
Using  $W_{I}=2.45 I-1.45 V$ %$W_{I}=2.41 I-1.41 V$ 
(for $R_V=3.3$ adopted in this study) and
the calibration % $M_W = -3.277 \log P - 2.69$ 
$M_W = -3.255 \log P - 2.644$ (based on $(m-M)_{0,LMC}=18.50$) \citep{uda99a,fre01},
we obtain from 21 Cepheids with classes 0, 1 and 2, 
 $(m-M)_0 = 24.52$ with $\sigma$(fit)=0.17, %$(m-M)_0 = 24.63$ with $\sigma$(fit)=0.17, 
which is smaller than the $\sigma$(fit)=0.25 from the the $I- \log P$ relation,
as shown in Figure 6.
From this we derive a distance modulus 
$(m-M)_0=24.52\pm0.15$(random)$\pm0.11$(systematic).
If we use 18 Cepheids with classes 0 and 1, 
we obtain  $(m-M)_0=24.55\pm0.15$(random)$\pm0.11$(systematic).
%$(m-M)_0=24.57\pm0.14$(random)$\pm0.10$(systematic).

We also tried to use the DIRECT mean magnitudes of the Cepheids corrected for
the difference between our the photometry and the DIRECT photometry.
The last two columns in Table 1 list the differences between  
the DIERCT magnitudes of the Cepheids at the same phase as the $HST$ data 
and the mean magnitudes.
Using these corrected mean magnitudes of the Cepheids, we 
obtain very similar results for the distance estimates to above (the difference in 
the distance modulus is only 0.02). 
Finally we adopt $(m-M)_0=24.52\pm0.14$(random)$\pm0.13$(systematic)
as the Cepheid distance to M33 before the metallicity effect correction.

\section{DISCUSSION AND SUMMARY}

  We have determined the distance to M33 using the single epoch $I$-band observations of
Cepheids based on the $HST/WFPC2$ images of five fields.
Combining the $HST$ $I$-band photometry and the periods determined from the ground-based
observations (DIRECT) for 21 Cepheids ($\log P >0.8$) 
with the best data in our sample of 
32 Cepheids, we derive a distance modulus of
$(m-M)_0=24.52\pm0.14$(random)$\pm0.13$(systematic).
%$(m-M)_0=24.41\pm0.13$(random)$\pm0.11$(systematic).
Using the Wesenheit $W_{I}$ quantity, an extinction-free parameter,
we derive a very similar value, 
$(m-M)_0=24.52\pm0.15$(random)$\pm0.11$(systematic).
These results are in good agreement with those based on 
the multi-epoch ground-based $BVRI$
observations of 11 bright Cepheids in M33,  
$(m-M)_0 = 24.56\pm0.10$ (and $E(B-V)=0.20\pm0.04$) 
before metallicity correction by \citet{fre91} and \citet{fre01}.
%$(m-M)_0 = 24.64\pm0.09$ %and (and $E(B-V)=0.10\pm0.09$) by \citet{fre91}.

 These Cepheid distances are somewhat smaller than those derived recently
using the tip of the red giant branch and red clump of the M33 field
stellar population. 
Using the same set of $HST$ data for field stars as we have used for the 
Cepheids, \citet{kim01} have determined the distance to M33,
obtaining $(m-M)_0=24.81\pm 0.04$(random)$^{+0.15}_{-0.11}$(systematic)
%$^0.14$ %(statistical only)
from the tip of the red giant branch (TRGB), and
$(m-M)_0=24.80\pm 0.04$(random)$\pm0.03$(systematic)
 from the mean magnitudes of the red clump (RC).
Note also that \citet{sar00} found $(m-M)_0 = 24.84 \pm 0.16$ from 
the inferred location of the RR Lyraes in two M33 halo globulars and
$24.81 \pm 0.24$ from the red clump of 7 halo clusters.

These TRGB distances and RC distances \citep{sar00,kim01} were derived adopting 
a foreground reddening of only $E(B-V)=0.04$.
The RGB and RC stars are old so that the M33 internal reddening for these stars 
is considered to be negligible.
On the other hand, Cepheid distances are derived adopting a total
reddening of $E(B-V)=0.20\pm0.04$ given by \citet{fre01}. 
%$E(B-V)=0.10$ given by \citet{fre91}. 
%The error of the total reddening for the Cepheids determined by \citep{fre91} 
%is as large as 0.09.
If only the foreground reddening is adopted for the Cepheids used in this study, 
the differences between the Cepheid distances and the TRGB and RC distances 
become much smaller (by 0.3 mag).
Since there is a large difference, $dE(B-V)=0.1$, in the reddening estimates
 based on the same data used by \citet{fre91} and \citet{fre01}, 
the uncertainty in the reddening must be larger than the quoted error
in \citet{fre01}, 0.04.
We strongly urge better determination of 
the reddening of the Cepheids in M33 in the future.

Note also that the metallicity effect in the Cepheid distance determination 
has been controversial
(see, e.g., \citet{sas97,koc97,ken98} and \citet{all01}). 
However, in the case of M33, 
the error due to metallicity differences is estimated to be negligible
because the mean metallicity of the disk components in M33 
is known to be very similar to or slightly more metal-rich
 than that of the LMC 
on which the calibration of the P-L relation used in this study is based.
While \citet{van00} lists $12+\log[O/H]=8.37\pm0.09$ for the LMC \citep{kur98}
and  $12+\log[O/H]=8.4\pm 0.15$ for M33 \citep{mas98},
\citet{fer00} and \citet{fre01} use $12+\log[O/H]=8.5\pm0.08$ for the LMC \citep{pag78}
and  $12+\log[O/H]=8.82\pm 0.15$ for M33\citep{zar94}. 
These values lead to a metallicity correction in the distance modulus to M33,
from $\delta(m-M)_0= 0.01$ to $0.06$, if the relation between the distance modulus
and metallicity adopted by \citet{fre01}, 
$\delta (m-M)_0 / \delta Z = -0.2\pm0.2$ mag dex$^{-1}$, is used.
However, we stress that the metallicity dependence of Cepheid luminosities
is very uncertain at the present time.

In closing, it is important to reiterate that the fitting error for 
the $I-\log P$ relation of 21 Cepheids based
on the single epoch $HST$ observations in this study is $\sigma$(fit) = 0.25.
%from which the error for the distance modulus is
%derived  to be $\sigma({\rm fit})/sqrt{N} = 0.25/sqrt{20} = 0.05$.
It is {\it impressive} that this fitting error is very similar to that 
for the mean $I$-band magnitudes
of 11 brighter Cepheids based on multi-epoch ground-based observations 
given by \citet{fre01},  $\sigma$(fit) = 0.27. % $\sigma$(fit) = 0.29.
This confirms that single epoch $I$-band observations of Cepheids 
using $HST$ is a very efficient way to determine accurate distances,
if the reddening of individual Cepheids can be accurately determined.
As a result, we would advocate the following strategy
for determining Cepheid distances to nearby galaxies:
First, search for Cepheids and determine the periods of the Cepheids using
small to mid-size ground-based telescopes (e.g. the DIRECT project);
Second, obtain $I$-band photometry of a large sample of  Cepheids at a single 
epoch (or a few epochs) using $HST$. 
Finally, determine the distance using the $I- \log P$ relation.
An excellent alternative is to
obtain K-band photometry of selected Cepheids.
 The very low K-band amplitude and low reddening make this a very promising way
to get accurate Cepheid distances to nearby galaxies, especially
those where reddening effects are important.

\acknowledgements
The authors are grateful to the referee, K. Stanek, for useful comments.
M.G.L. is supported in part by the MOST/KISTEP International 
Collaboration Research Program (1-99-009). 
M.G.L. is grateful to the Astronomy Group
at the University of Concepcion for the warm hospitality during his stay for this work.
A.S. has benefitted from financial support from NSF CAREER grant No. AST-0094048.
D.G. and W.G. acknowledges financial support for this project received from CONICYT 
through Fondecyt grant 8000002.

%\input{table1.tex}

%\documentclass[12pt]{article}
%\documentclass[preprint2]{aastex}
%\documentclass[aj_pt4]{aastex}
%\documentclass{aastex}

%\newcommand{\vdag}{(v)^\dagger}
%\newcommand{\myemail}{mskim@astro.snu.ac.kr}

%\slugcomment{\today }

%\shorttitle{Distances to M33} \shortauthors{Lee et al.}

%\begin{document}

\begin{deluxetable}{rcrccccccrr}
\tablecaption{$VI$ Photometry of Cepheids in the HST fields of M33\label{tbl-1}} \tablewidth{0pt}
\tablecolumns{6}\tablehead{\colhead{Name\tablenotemark{a} }  &
\colhead{DIRECT Name\tablenotemark{b}}& 
\colhead{P[days]\tablenotemark{b}} &
\colhead{Phase\tablenotemark{c}} &
\colhead{$V$\tablenotemark{d}} &
\colhead{$V-I$\tablenotemark{d}} &
\colhead{Class\tablenotemark{e}} &
%\colhead{$<B>$\tablenotemark{f}} &
\colhead{$<V>$\tablenotemark{f}} &
\colhead{$<I>$\tablenotemark{f}} &
\colhead{$\Delta V$\tablenotemark{e}} &
\colhead{$\Delta I$\tablenotemark{e}} \\}
\startdata
   C5 & D33J013349.8+304737.0 &   4.19 &    0.34 &   22.00 &    0.89 &  2 &      22.58 &   21.01 &    0.36 &    0.02 \\ 
  C12 & D33J013359.4+304214.2 &   4.78 &    0.35 &   21.47 &    1.03 &  2 &      21.67 &   20.85 &    0.19 &    0.24 \\ 
  C24 & D33J013408.6+303754.8 &   5.32 &    0.89 &   21.08 &    0.65 &  0 &     21.23 &      -- &   --0.27 &      -- \\ 
  C27 & D33J013405.5+304133.3 &   5.37 &    0.65 &   21.61 &    0.84 &  2 &     21.97 &      -- &   --0.22 &      -- \\ 
  C29 & D33J013359.5+303846.8 &   5.45 &    0.63 &   21.04 &    0.91 &  1 &     21.24 &      -- &   --0.01 &      -- \\ 
  C35 & D33J013351.2+303001.0 &   5.60 &    0.29 &   20.64 &    0.69 &  1 &    20.58 &   19.63 &   --0.07 &   --0.03 \\ 
  C43 & D33J013350.6+304734.9 &   5.74 &    0.28 &   21.55 &    0.85 &  1 &      21.60 &   20.74 &   --0.07 &   --0.12 \\ 
  C49 & D33J013407.9+303831.6 &   5.90 &    0.82 &   21.93 &    1.14 &  4 &     20.31 &   19.82 &   --0.15 &   --0.12 \\ 
  C55 & D33J013349.6+304744.7 &   6.00 &    0.58 &   21.65 &    0.90 &  2 &      21.92 &   20.52 &   --0.21 &   --0.26 \\ 
  C62 & D33J013405.6+304120.5 &   6.12 &    0.96 &   21.25 &    0.84 &  1 &    21.58 &   20.35 &   --0.24 &   --0.21 \\ 
  C66 & D33J013349.4+304701.9 &   6.35 &    0.87 &   21.39 &    0.89 &  1 &     21.43 &      -- &    0.18 &      -- \\ 
  C68 & D33J013349.4+303009.4 &   6.78 &    0.15 &   21.39 &    0.92 &  0 &     21.12 &   20.40 &    0.39 &    0.29 \\ 
  C71 & D33J013406.4+304003.7 &   6.93 &    0.36 &   20.65 &    0.68 &  0 &      20.57 &   19.59 &   --0.23 &   --0.07 \\ 
  C72 & D33J013428.3+303900.4 &   6.99 &    0.76 &   21.06 &    0.72 &  0 &     21.24 &   20.41 &   --0.16 &   --0.12 \\ 
  C83 & D33J013347.9+302943.6 &   7.63 &    0.02 &   21.12 &    0.85 &  1 &      21.33 &   19.94 &   --0.05 &   --0.07 \\ 
 C100 & D33J013402.4+303831.8 &   8.33 &    0.96 &   21.10 &    0.85 &  2 &     20.80 &   20.17 &   --0.01 &    0.03 \\ 
 C104 & D33J013406.6+303816.8 &   8.58 &    0.75 &   21.04 &    0.86 &  0 &    20.29 &      -- &    0.18 &      -- \\ 
 C111 & D33J013346.3+302908.9 &   9.12 &    0.55 &   20.72 &    0.77 &  1 &     20.59 &   19.83 &   --0.25 &   --0.12 \\ 
 C118 & D33J013350.8+304715.5 &   9.72 &    0.94 &   20.70 &    0.86 &  1 &    20.82 &   20.13 &   --0.02 &   --0.11 \\ 
 C121 & D33J013408.8+303946.5 &  10.11 &    0.73 &   20.77 &    0.84 &  0 &     20.56 &   19.52 &    0.02 &    0.03 \\ 
 C142 & D33J013357.4+304113.9 &  11.62 &    0.61 &   21.33 &    1.21 &  0 &    20.34 &   19.35 &    0.23 &    0.20 \\ 
 C150 & D33J013346.6+304821.8 &  12.35 &    0.32 &   20.55 &    1.73 &  3 &     20.22 &   18.49 &    0.14 &    0.06 \\ 
 C156 & D33J013350.0+303014.9 &  12.97 &    0.31 &   20.51 &    0.84 &  0 &    20.72 &   19.39 &   --0.36 &   --0.14 \\ 
 C157 & D33J013406.1+303734.0 &  13.02 &    0.45 &   21.32 &    1.36 &  2 &    21.06 &   19.71 &    0.16 &    0.08 \\ 
 C158 & D33J013402.8+304145.7 &  13.04 &    0.33 &   20.41 &    0.96 &  0 &    20.02 &   18.91 &    0.09 &   --0.00 \\ 
 C160 & D33J013359.9+303910.3 &  13.24 &    0.38 &   20.30 &    0.97 &  0 &     20.67 &   19.33 &   --0.24 &    0.00 \\ 
 C163 & D33J013408.1+303931.9 &  13.32 &    0.35 &   21.16 &    1.07 &  0 &    20.41 &   19.83 &    0.06 &    0.16 \\ 
 C174 & D33J013405.9+303928.9 &  14.59 &    0.97 &   20.45 &    0.98 &  0 &     20.01 &   19.32 &    0.37 &    0.19 \\ 
 C204 & D33J013346.6+304645.9 &  18.81 &    0.04 &   20.13 &    0.95 &  0 &    19.79 &   18.95 &    0.32 &    0.06 \\ 
 C205 & D33J013406.8+303940.2 &  18.89 &    0.07 &   20.54 &    1.08 &  0 &    19.84 &   18.96 &   --0.09 &    0.00 \\ 
 C212 & D33J013401.7+303923.1 &  21.67 &    0.67 &   20.28 &    1.24 &  2 &   19.89 &   18.80 &    0.03 &   --0.05 \\ 
 C224 & D33J013350.6+304754.7 &  26.48 &    0.47 &   20.51 &    1.30 &  0 &    20.02 &   18.73 &    0.22 &    0.10 \\ 
\enddata
\tablenotetext{a}{Our IDs for the Cepheids.}
\tablenotetext{b}{Periods from the DIRECT Project \citep{mac01}.}
\tablenotetext{c}{Phases of the Cepheids in the $HST$ photometry.}
\tablenotetext{d}{Single epoch $HST$ photometry of this study. Formal errors are smaller than 0.02 mag.}
\tablenotetext{e}{Classes depending on the quality of the DIRECT light curves of Cepheids: 0=very good, 1=good, 2=fair,
3=unusually red color for Cepheids, and 4=ambiguous identification.}
\tablenotetext{f}{Mean magnitudes of Cepheids from the DIRECT photometry.}
\tablenotetext{e}{Differences between  
the DIRECT magnitudes of Cepheids at the same phase
as the $HST$ data and the mean magnitudes.}
\end{deluxetable}
\clearpage

\begin{deluxetable}{lc}
\tablecaption{Error budget for the Cepheid distance modulus to M33\label{tbl-2}} \tablewidth{0pt}
\tablecolumns{2}\tablehead{\colhead{Source of Error}&\colhead{Error} } %\\}
\startdata
Cepheid PL calibration & \\
  ~~A. LMC true modulus & 0.10 \\
  ~~B. $I$-band PL zero point & 0.01 \\
  ~~S1. LMC PL systematic uncertainty & 0.10 \\
  ~~S2. Metallicity effect (LMC--M33) & 0.04 \\
HST photometric calibration & \\
  ~~D. HST $I$-band zero point & 0.05\\
  ~~DW. HST $V$-band zero point & 0.05\\
  ~~R1. HST calibration uncertainty & 0.12 \\
  ~~RW1. HST calibration uncertainty & 0.14 \\
M33 distance modulus & \\
  ~~F. M33 $I$-band PL fitting & 0.05 \\
  ~~FW. M33 $W-\log P$ PW fitting & 0.04 \\
  ~~H. Finite strip width and random phase data & 0.07 \\
  ~~R2. M33 distance modulus uncertainty & 0.07\\
  ~~RW2. M33 distance modulus uncertainty & 0.04\\
  ~~S3. Total $I$-band extinction uncertainty & 0.07 \\ 
Total uncertainty\tablenotemark{a} & \\
~~R. Random errors for PL and PW & 0.14, 0.15 \\
~~S. Systematic errors for PL and PW & 0.13, 0.11 \\ 
\enddata
\tablenotetext{a}{ $R(PL)=\sqrt{R1^2+R2^2}$,
                   $R(PW)=\sqrt{RW1^2+RW2^2}$,
                   $S(PL)=\sqrt{S1^2+S2^2+S3^2}$, and
                   $S(PW)=\sqrt{S1^2+S2^2}$.}
\end{deluxetable}

%\end{document}

\clearpage
\begin{figure}
\epsscale{0.5}
\plotone{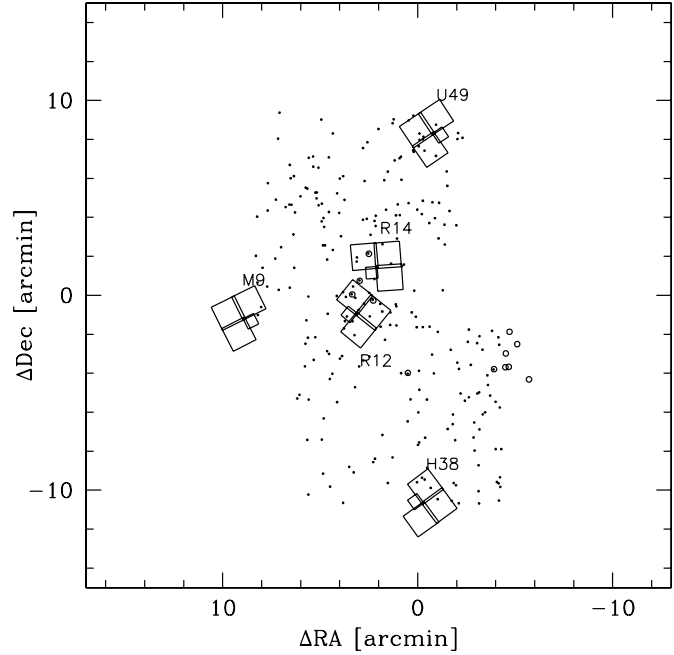}
\figcaption[fig1.ps]{A finding chart for Cepheids in M33.
The dots represent the DIRECT Cepheids, and the open circles represent the
Cepheids used in \citet{fre91}. The squares represent the $HST$ fields centered on
the globular clusters.
\label{fig1}}
\end{figure}

\begin{figure}
\plotone{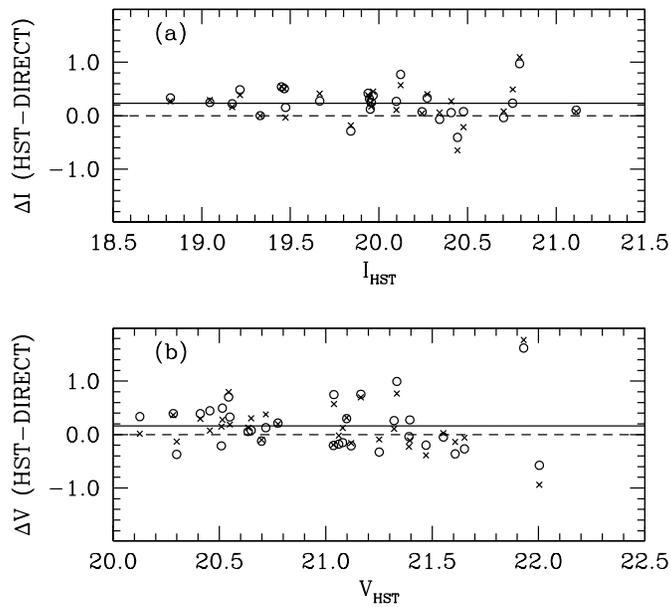}
\figcaption[fig2.ps]{Comparisons of the $HST$ photometry and DIRECT photometry of
Cepheids in the $I$ (a) and $V$ (b) bands. The crosses represent the difference
between the $HST$ photometry and the DIRECT magnitudes at the same phase, and
the open circles
 represent the difference between the $HST$ photometry and the DIRECT mean magnitudes of the Cepheids. 
The solid lines represent the mean value for
the differences between the $HST$ photometry and the DIRECT mean
magnitudes.
\label{fig2}}
\end{figure}

\begin{figure}
\plotone{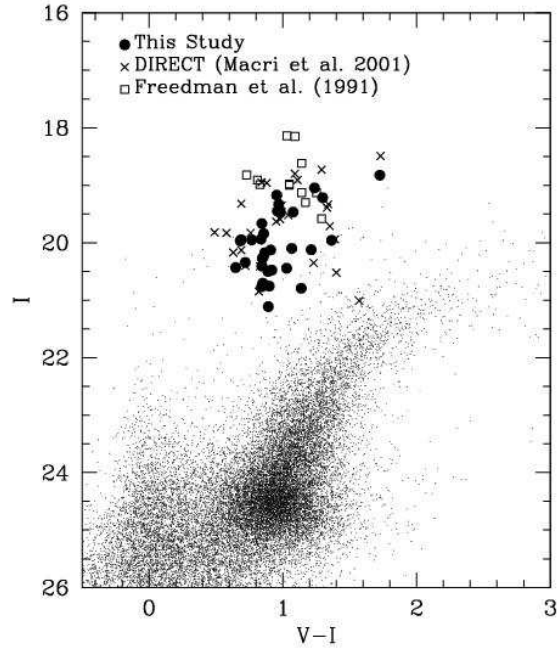}
\figcaption[fig3.ps]{Color-magnitude diagram of Cepheids in M33.
The filled circles represent our single epoch photometry of the Cepheids identified in
the $HST$ images,
and the crosses represent the mean magnitudes and colors of the same Cepheids given by
the DIRECT project, and
the open squares represent the mean magnitudes and colors of bright Cepheids given by \citet{fre91}.
The reddest Cepheid is C150.
The dots represent the field stars in the WF3 chip of
the H38-region, as an example.
\label{fig3}}
\end{figure}

\begin{figure}
\plotone{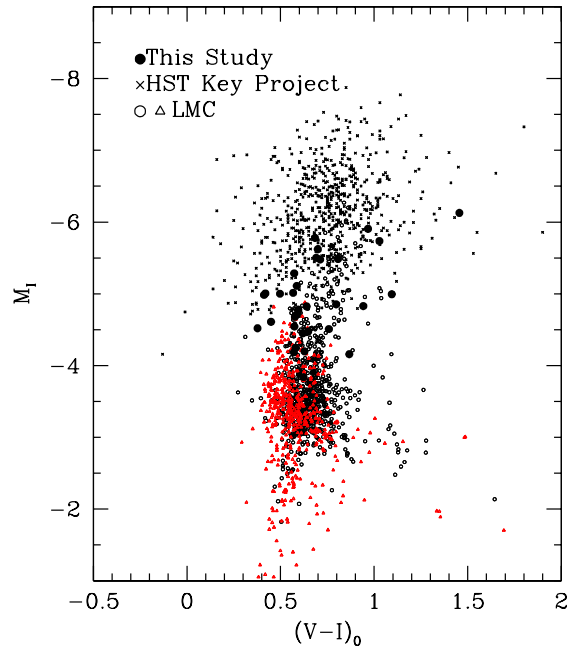}
\figcaption[fig4.ps]{Color-magnitude diagram of Cepheids 
in M33 (filled circles), LMC (open circles and triangles), 
and $HST$ $H_0$ Key Project galaxies (crosses).
Extinction values and distance moduli given by \citet{fre01} are adjusted for
M33 and eighteen $HST$ $H_0$ Key Project galaxies. 
The open circles represent the fundamental mode Cepheids and 
the open triangles represents the first overtone Cepheids in the LMC
from the OGLE data \citep{uda99b}. The extinction $E(B-V)=0.1$
and the distance modulus $(m-M)_{0}=18.5$ for the LMC are adopted.
\label{fig4}}
\end{figure}

\begin{figure}
\plotone{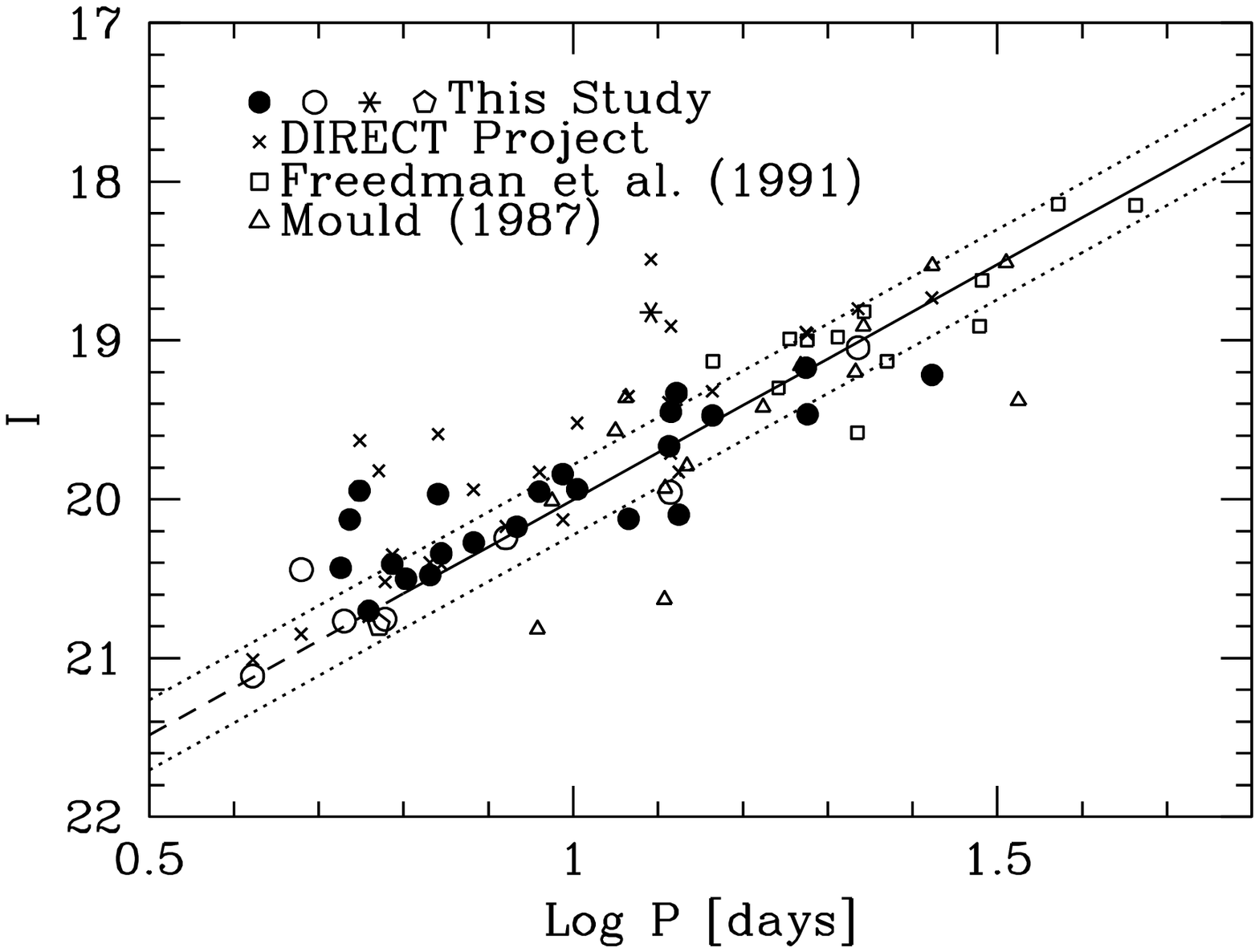}
\figcaption[fig5.ps]{
Distance estimation using the $I-\log P$ relation 
(given by \citet{fre01}) for the Cepheids in M33.
The solid line represents a linear fit to the data of
the Cepheids with $\log P > 0.8$ and classes 0 and 1 
obtained in this study (filled circles), 
and the dotted lines represent 2$\sigma$ (=0.22) excursions  
from the $I - \log P $ relation for the LMC.
The open circles, asterisks and open pentagons represent, respectively, 
the Cepheids with class 2, 3 and 4 in this study, and the crosses represent  
the mean magnitudes of the same Cepheids given by the DIRECT project \citep{mac01}.
The open squares  represent
 the mean magnitudes of  bright Cepheids given by \citet{fre91}, 
and the open triangles  represent the single epoch magnitudes of 
other Cepheids given by \citet{mou87}. 
\label{fig5}}
\end{figure}

\begin{figure}
\plotone{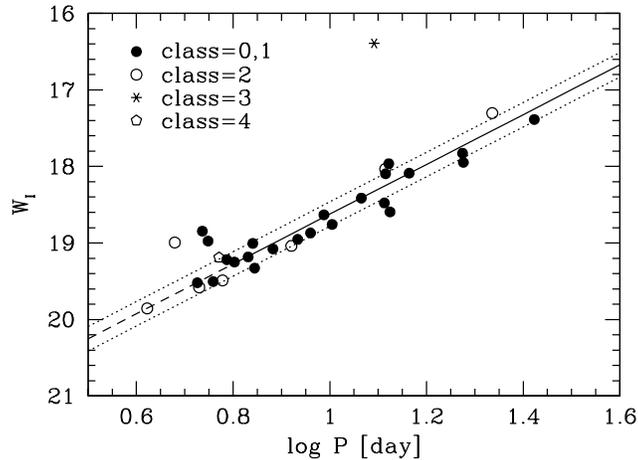}
\figcaption[fig6.ps]{
Distance estimation using the $W_{I}-\log P$ relation
for the Cepheids in M33.
The solid line represents a linear fit to the data of
the Cepheids with $\log P > 0.8$ and classes 0 and 1 
obtained in this study (filled circles),   and the dotted lines
represent 2$\sigma(=0.16)$ from the calibration.
The open circles, asterisks and open pentagons represent, respectively, 
the Cepheids with classes 2, 3 and 4 in this study.
\label{fig6}}
\end{figure}

\clearpage
\end{document}